\documentclass[twocolumn,aps,amsmath,nofootinbib,superscriptaddress]{revtex4}
\usepackage{psfig}
\usepackage{graphicx}
\usepackage[latin1]{inputenc}

\usepackage{color}

\definecolor{joerg}{rgb}{1.0,0.0,0.0}

\begin{document}

\title{Determination of the Fundamental Scale of Gravity and \\ the Number
of Space-time Dimensions from High Energetic Particle Interactions}

\author{J. Ruppert}

\address{{}Department of Physics, Duke University, Science Drive, Box 90305 Durham, NC 27708, USA}

\author{C.~Rahmede}

\address{SISSA, Scuola Internazionale Superiore di Studi Avanzati,\\
               via Beirut 4, 34014 Trieste, Italy}

\address{{}Institut f\"ur Theoretische Physik, J. W. Goethe Universit\"at, \\
Robert-Mayer-Str. 10, 60054 Frankfurt am Main, Germany}

\author{M. Bleicher}

\address{{}Institut f\"ur Theoretische Physik, J. W. Goethe Universit\"at, \\
Robert-Mayer-Str. 10, 60054 Frankfurt am Main, Germany}

\maketitle
Within the ADD-model, we elaborate an idea by Vacavant and Hinchliffe \cite{Vacavant:sd} and show quantitatively
how to determine the fundamental scale of TeV-gravity and the number of compactified extra dimensions from data at LHC.
We demonstrate that the ADD-model leads to strong correlations between the missing $E_T$
in gravitons at different center of mass energies. This correlation puts strong constraints
on this model for extra dimensions, if probed at $\sqrt{s}=5.5$~TeV  and $\sqrt{s}=14$~TeV  at LHC.\vspace*{1cm}\\

Recently string theory motivated models with additional space-time dimensions have moved into the center of attention in high energy physics. Depending on the size of the extra dimensions and the geometry of space-time three different kinds of extra dimensional models are usually discussed: the model of universal extra dimensions \cite{Antoniadis:1998ig} which allows all particles to propagate into the new dimensions, the model of Randall and Sundrum (RS) \cite{RandallSundrum} with one ''gravity-only'' extra dimension and the ADD-model \cite{Arkani-Hamed:1998rs,Arkani-Hamed:1998nn}
 with many ''gravity-only'' extra dimensions. Especially the RS- and ADD-models allow the introduction of a new fundamental scale $M_D$ of gravity in the TeV range. This drastic increase of the coupling strength of gravity on small scales compared to the Planck scale
results in a vast amount of potentially observable effects:

\begin{itemize}
\item
Black hole production in colliders and ultra high energetic cosmic rays (UHECR) \cite{bf,Giddings:2001ih,Giddings:2001bu,Ringwald:2001vk,Hossenfelder:2001dn,Emparan:2001kf,Bleicher:2001kh,Hofmann:xd,Cavaglia:2002si,Cavaglia:2003qk,Hossenfelder:2003dy},
\item
increased neutrino cross sections in UHECR interactions \cite{Nussinov:2001pu,Jain:2000vg,Anchordoqui:2002hs,Han:2004kq},
\item
virtual graviton exchange processes\cite{Giudice:1998ck,Han:1998sg,Atwood:1999cy,Dvergsnes:2002nc,Buanes:2004ya},
\item
direct graviton production as Kaluza-Klein resonances \cite{Giudice:1998ck,Han:1998sg,Mirabelli:1998rt,Antoniadis:2000vd,Vacavant:sd,Atwood:1999pn,Atwood:1999zg,Ahern:2000jn,Rizzo:1999pc,Rizzo:2001ag,Cullen:2000ef}.
\end{itemize}

Present limits on the new fundamental scale and the size and number of extra dimensions have been obtained from: direct measurements of the gravitational inverse square law \cite{Adelberger}, hadron-hadron interactions at Tevatron \cite{Abbott:2000zb,Acosta:2003tz}, modifications of cosmic ray cross sections \cite{Emparan:2001kf,Anchordoqui:2001cg,Kazanas:2001ep,Sigl:2002bb} and supernova explosions and cooling \cite{Cullen:1999hc,Hannestad:2003yd}.

While supernova cooling gives a very tight constraint of $M_D>500$~TeV for $\delta=2$, more than two extra dimensions lead to constraints of $M_D$ of the order of a TeV. The Tevatron data constraints $M_D$ to be of the order or above $1$ TeV.

In this letter we elaborate on an idea by Vacavant and Hinchliffe \cite{Vacavant:sd}. They discuss qualitatively how to 
determine the number of extra dimensions from the ratio of cross sections for missing transverse energy at different 
center-of-mass (CMS) energies. In addition to their analysis we focus on quantitative predictions and the strong 
correlations of the cross sections for graviton production at LHC within the ADD-model. 

Our presentation of the cross section for  missing energy at 
$\sqrt{s}=5.5$~TeV and $\sqrt{s}=14$~TeV pp collisions allows to directly read off both the $M_D$ and $\delta$ 
values from  the experimentally measured  missing energy cross sections from graviton production.
Furthermore, we show that the ADD-model predicts very strong correlations of these cross sections for different CMS energies
providing a crucial test of the ADD-scenario.

For the calculations employed here we use the leading order parton~$+$~parton $\rightarrow$ graviton~$+$~parton cross sections given in \cite{Giudice:1998ck}. The reader should be aware that the use of leading order cross sections can only be justified up to parton-parton center of mass energies of $\sqrt{\hat s}\le 6 M_D$. Thus, for $M_D\le 2$~TeV and $\sqrt{s_{\rm pp}} =14$~TeV the present results might achieve corrections.  
The differential cross section for the production of a jet and a graviton in pp interactions is then obtained by folding the two particle cross sections with the parton distribution functions $f_{i}$ (here we use CTEQ6 \cite{Pumplin:2002vw,Stump:2003yu}):
 \begin{eqnarray}\label{Jetcrosssection}
&&\frac{{\rm d^3} \sigma}{{\rm d} y {\rm d} p_T{\rm d} m}
 \left(AB \rightarrow jet + G\right)=\nonumber\\
&&2 p_T \sum_{\rm partons}
\int_{x_{\rm min}}^1 {\rm d} x_a 
  \frac {x_a x_b}{x_a - \frac {m_T}
  {\sqrt s}\, e^y} \nonumber\\
 &&f_{a}\left(x_a,Q^2\right) f_{b}\left(x_b,Q^2\right)
  \frac{\rm d^2 \sigma}{{\rm d}\hat{t}{\rm d} m}\left(a\,b \rightarrow c\,G\right)
\end{eqnarray}
\noindent
with the transverse graviton mass $m_T$, the rapidity $y$, $Q^2=2 \hat s  \hat t \hat u/(\hat s^2+\hat t^2+\hat u^2)$, and  
\begin{equation}\label{gravcs}
    \frac{\rm{d}^2 \sigma}{\rm{d}t\rm{d}m}=S_{\delta-1}\frac{M_{Pl}^2}{M_D^{2+\delta}}m^{\delta-1}\frac{\rm{d} \sigma_m}{\rm{d}t}\quad, 
\end{equation}
$m$ being the mass of the graviton and $S_{\delta-1}$ the surface of the $
\delta$-unit
sphere.
$d\sigma_m/dt$ is the elementary cross section for the production of a graviton of mass $m$ \cite{Giudice:1998ck}. 
It is interesting to note that the $1/M_{Pl}^2$ in the transition matrix is cancelled by the 
phase space factor resulting in an enhanced cross section $\propto 1/M^{2+\delta}_D$.

Due to their small interaction cross section with standard model particles and their long lifetimes gravitons escape the detector region without a signal. Thus, gravitons will be observed indirectly by missing transverse energy.

Here we quantify the energy loss by demanding a minimum missing transverse energy  $E_{T,min}$ in the mid-rapidity range ($-3\le y\le 3$):
\begin{eqnarray}
\label{intsigma}
&&\left. \sigma\left(AB \rightarrow jet + G\right)\right|_{E_{T,min}}=\\ \nonumber
&&\int_{-3}^3{\rm d} y
\int_{E_{T,min}}^{\infty}\!{\rm d}E_T
\int_0^{\sqrt s/2}{\rm d} m 
\frac{{\rm d} \sigma(AB \rightarrow jet + G)}{{\rm d} y {\rm d} p_T{\rm d} m}
\end{eqnarray}
\noindent

\begin{figure}
 \par\resizebox*{!}{0.38\textheight}{\includegraphics{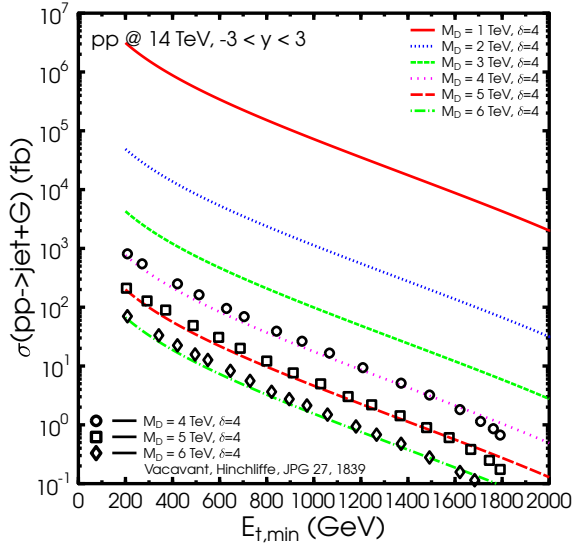}} \par{}
%\vspace*{.8cm}
\caption{\label{fig1} Integrated cross section for missing $E_T>E_{T,min}$ in pp-collisions at $\sqrt{s}=14$~TeV for four extra dimensions and different fundamental scales $M_D$. Lines denote our calculation, symbols show calculations by \cite{Vacavant:sd}.}
\end{figure}

In Fig. \ref{fig1} we show the integrated cross section for missing energy as given by 
Eq. \ref{intsigma} for four extra dimensions. The lines (from top to bottom) show the results for different values of the fundamental scale $M_D$ from $1$ to $6$~TeV.
As a check we compare to  \cite{Vacavant:sd} (symbols).

Let us now focus on how to extract the fundamental scale and the number of space time dimensions in the ADD-model from data.
The cross section for a mono-jet- and missing energy event depends on $M_D$ and $\delta$, however, information on the cross-section at only one CM-energy leads to a set of different possible $\delta$ and $M_D$. Here, we suggest to combine more than one cross section measurements at different CMS energies. This allows to determine the $\delta$ and $M_D$ value, uniquely.

To be specific we chose for the following analysis pp collisions at $\sqrt{s}=14$~TeV and $\sqrt{s}=5.5$~TeV. If at $\sqrt{s}=5.5$~TeV only Pb+Pb data will be available, our proton-proton prediction could be scaled up by the number of binary collisions to the heavy system (neglecting shadowing corrections).\\  

\begin{figure}
 \par\resizebox*{!}{0.32\textheight}{\includegraphics{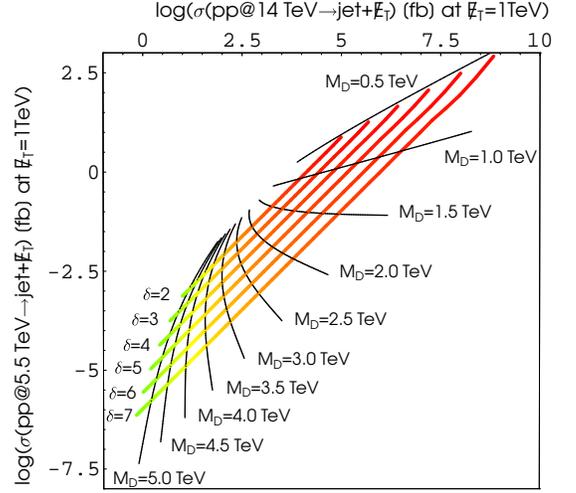}} \par{}
%\vspace*{.8cm}
\caption{\label{fig2} Combinations of cross section at $E_{T,min}=1$~TeV for pp collisions at $\sqrt{s}=5.5$~TeV (horizontal axis) and $\sqrt{s}=14$~TeV (vertical axis). The thick lines denote possible cross section combinations from the ADD-model for fixed values of $\delta$ and varying $M_D$. The thin lines indicate equi-$M_D$ values on the thick lines.}
\end{figure}

For both CMS energies we extract the $\delta$ and $M_D$ dependent cross sections at 
$E_{T,min}=1$~TeV. Fig. \ref{fig2} shows the extracted combinations of cross sections consistent with the ADD-model. The thick lines denote calculations for fixed $\delta$ and varying  $M_D$ while the thin lines indicate fixed $M_D$ values. From this correlation plot two qualitatively different conclusions can be drawn when data becomes available:

\begin{itemize}
\item
If the measurements are off the thick lines, the missing energy cannot be explained by graviton production in the ADD-model. 
\item
If the measurements are compatible with one of the thick lines, the missing energy can be attributed to graviton production in the ADD-scenario. Even more, the number of extra dimensions $\delta$ and the new fundamental scale $M_D$ can be directly extracted from Fig. \ref{fig2}.
\end{itemize}

In conclusion, within the ADD-model we predict strong correlations between the missing energies observed at different CMS energies at LHC. If the observed energy loss is in agreement with the present calculation it is possible to extract both the number of extra dimensions and the fundamental scale of gravity, uniquely, at the LHC.\\[1.2cm]

\section*{Acknowledgements}
This work was supported by GSI, DFG, BMBF.
JR thanks the Alexander von Humboldt foundation as a Feodor Lynen fellow.
We thank Sabine Hossenfelder for discussions.

\end{document}